\documentclass{aa}
\usepackage{graphicx,natbib}
\bibpunct{(}{)}{;}{a}{}{,}

\def\apj{ApJ}

\def\flux{erg s$^{-1}$ cm$^{-2}$}
\def\lum{erg s$^{-1}$}

\begin{document}

\sloppypar

%   \thesaurus{08     % A&A Section 6: Form. struct. and evolut. of stars
%              (02.01.2;  % Accretion, accretion disks
%               02.09.1;  % Instabilities
%              08.02.3;  % Stars:binaries:general
%              08.06.3;  % Stars:classification
%              08.14.1;  % Stars: neutron
%               13.25.3;  % X-rays: general
%               13.25.5)} % X-rays: stars
%
   \title{Resolving the Galactic X-ray background}

   \author{M.~Revnivtsev \inst{1,2} \and A. Vikhlinin \inst{3,2} \and S.~Sazonov \inst{1,2}}

   \offprints{mikej@mpa-garching.mpg.de}

   \institute{
              Max-Planck-Institute f\"ur Astrophysik,
              Karl-Schwarzschild-Str. 1, D-85740 Garching bei M\"unchen,
              Germany,
      \and
              Space Research Institute, Russian Academy of Sciences,
              Profsoyuznaya 84/32, 117997 Moscow, Russia
       \and
               Harvard-Smithsonian Center for Astrophysics, 60 Garden Street, Cambridge, MA 02138, USA
            }
  \date{}

        \authorrunning{Revnivtsev et al.}

        \abstract{We use Chandra deep observations of the Galactic
          Center (GC) region to improve the constraints on the
          unresolved fraction of the Galactic X-ray background (also
          known as the Galactic ridge X-ray emission). We emphasize the
          importance of correcting the measured source counts at low
          fluxes for bias associated with Poisson noise. We find that at
          distances of $2^\prime$--$4^\prime$ from Sgr A$^*$ at least
          $\sim 40$\% of the total X-ray emission in the energy band
          4--8~keV originates from point sources with luminosities
          $L_{\rm 2-10~keV}>10^{31}$ \lum. 
From a comparison of the source number-flux
          function in the GC region with the known luminosity function
          of faint X-ray sources in the Solar vicinity, we infer that
          Chandra has already resolved a large fraction of the
          cumulative contribution of cataclysmic variables to the total
          X-ray flux from the GC region. This
          comparison further indicates that most of the yet unresolved
          $\sim 60$\% of the X-ray flux from the GC region is likely
          produced by weak cataclysmic variables and coronally active stars with $L_{\rm
            2-10~keV}<10^{31}$ \lum. We conclude that the bulk of the
          Galactic X-ray background is produced by discrete
          sources.  \keywords{stars: binaries: general -- Galaxy: bulge
            -- Galaxy: disk -- X-rays: general -- X-rays: stars } }

   \maketitle

%
%________________________________________________________________

\section{Introduction}
One of the largest extended features of the X-ray sky is the Galactic
X-ray background, often referred to as the Galactic ridge X-ray emission
(GRXE hereafter), discovered in the late 1970's
\citep{cooke70,bleach72,worrall82}. The GRXE extends over more than 100
degrees along the Galactic plane but only a few degrees across it
\citep[e.g.,][]{warwick85,yamauchi93}.

The origin of the GRXE is the topic of a long-standing debate. The GRXE
spectrum resembles the optically thin emission of thermal plasma with a
temperature of 5--10~keV \cite[e.g.][]{koyama86,koyama89}. However, such
hot plasma cannot be gravitationally bound to the Galaxy, as suggested
by strong concentration of the GRXE towards the Galactic disk and bulge.
A number of models, still assuming a diffuse origin of the GRXE, were
proposed \cite[see e.g. the review by][]{tanaka02}, but none of them was
successful in explaining all of observed properties of the GRXE.

An alternative explanation of the GRXE, proposed soon after its
discovery, is that it is the superposition of weak Galactic X-ray
sources \citep[e.g.][]{worrall82,worrall83,koyama86,ottmann92,mukai93}.
However, until recently there remained a large uncertainty with regard
to the expected contributions of different classes of faint Galactic
X-ray sources to the GRXE.

In recent work of \cite{mikej06}, \cite{mikej06_67kev}, and
\cite{krivonos06}, based on X-ray observations from the RXTE and
INTEGRAL satellites, it was demonstrated that the GRXE closely traces
the near-infrared emission and consequently the stellar mass
distribution in the Galaxy. Furthermore, the observed GRXE to stellar
mass ratio is compatible with the X-ray luminosity function of
cataclysmic variables and coronally active stars in the vicinity of the
Sun, measured with RXTE and ROSAT \citep{sazonov06}. These findings
therefore suggest that the bulk of the GRXE could indeed be produced by
emission from descrete sources \citep{mikej06}.

A straightforward way to test the origin of the GRXE is to check which
fraction can be resolved into discrete sources in deep \emph{Chandra}
images. The first such study was by
\cite{ebisawa01,ebisawa05} who used a $\sim 200$~ks Chandra observation of a
Galactic plane region. They inferred that less than 15\% of the GRXE in
that region could be resolved into point sources with fluxes higher than
(3--5)$\times 10^{-15}$ \flux.

\begin{figure}
\includegraphics[width=\columnwidth]{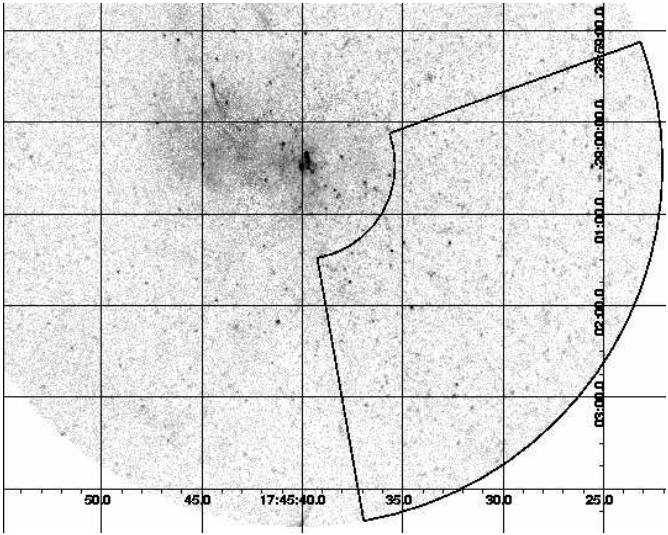}
\caption{Chandra image of the Galactic Center region in the energy band 4--8
keV (efficiency corrected). The area of our study is outlined by the
solid line.} 
\label{image}
\end{figure}

Perhaps the best place for studying the Galactic X-ray background is the
Galactic Center (GC) region, because: 1) the GRXE spectrum in this
region is fairly typical \citep[e.g.,][]{tanaka02}, 2) the GRXE
intensity is so high that the contribution of the extragalactic X-ray
and the instrumental backgrounds is very low, and 3) the high
concentration of X-ray sources near the GC implies that most of them are
within the Galactic nuclear stellar cluster and thus at the known
distance from the Sun; this allows one to easily convert the source
$\log N - \log S$ relation to the intrinsic luminosity function. A
disadvantage of the GC region for the GRXE studies is the presence of a
large number of supernova remnants and different types of non-thermal
phenomena \citep{muno04,park04}.

\emph{Chandra} has accumulated a long exposure time in the several
arcmin$^2$ near the GC. Previous analyses of the combined $\sim
600$~ksec worth of data \citep{muno03,muno04,park04} have shown that
$\sim20$--30\% of the total X-ray flux can be resolved into point
sources in a sub-region that is maximally free from supernova remnants
\citep[region ``Close'' in][]{muno04}. Since these works, an additional
330~ksec of GC observations have become available, which allows one to
go deeper in resolving the Galactic X-ray background. In this paper, we
analyze all publicly available \emph{Chandra} data in the GC region in
an attempt to resolve as much of the GXRE as possible.

\begin{figure}
\includegraphics[width=\columnwidth]{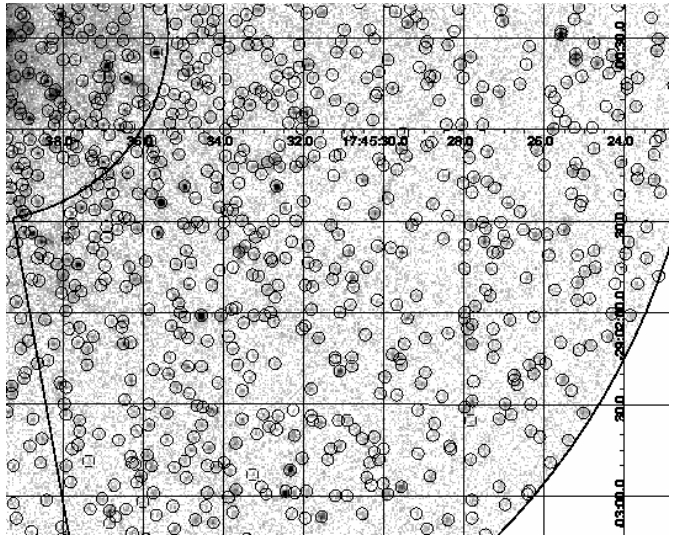}
\caption{Part of the image of the Galactic Center region in the energy
band 4--8 keV, with the detected sources shown by circles (note that the
source localizations are actually much more accurate than $\sim
2.5$~arcsec, the circle radius).} 
\label{image_sources}
\end{figure}

\section{Data reduction}
\label{sec:data-reduction}
We used \emph{Chandra} ACIS-I observations of the GC region with the aim
point close to Sgr A$^*$ (observation ID \#945, 1561, 2282, 2284, 2287,
2291, 2293, 2943, 2951, 2952, 2953, 2954, 3392, 3393, 3663, 3665, 4500,
4683, 4684, 5360, and 6113). The \emph{Chandra} data were reduced
following a standard procedure fully described in
\cite{2005ApJ...628..655V}. The only difference is that the detector
background was modeled using the stowed dataset
(http://cxc.harvard.edu/contrib/maxim/stowed). The total clean exposure
time is 918~ks.  Figure~\ref{image} shows the combined image in the
4--8~keV energy band 4--8 keV.

The GC region is known to be rich in supernova remnants \citep[see
e.g.][]{park04,muno04}. In order to minimize the contribution of X-ray
emission from plasmas heated by the supernova remnants, in our analysis
we: 1) considered only a sub-region to south-west of 
the GC (see Fig. \ref{image}), subtending 9.77 arcmin$^2$, 
where the contribution of low-temperature plasmas is known to be small
from the weakness of low-energy X-ray lines \citep[see e.g.][]{park04}
and 2) used only the energy band 4--8 keV.

\begin{figure}
\includegraphics[width=\columnwidth]{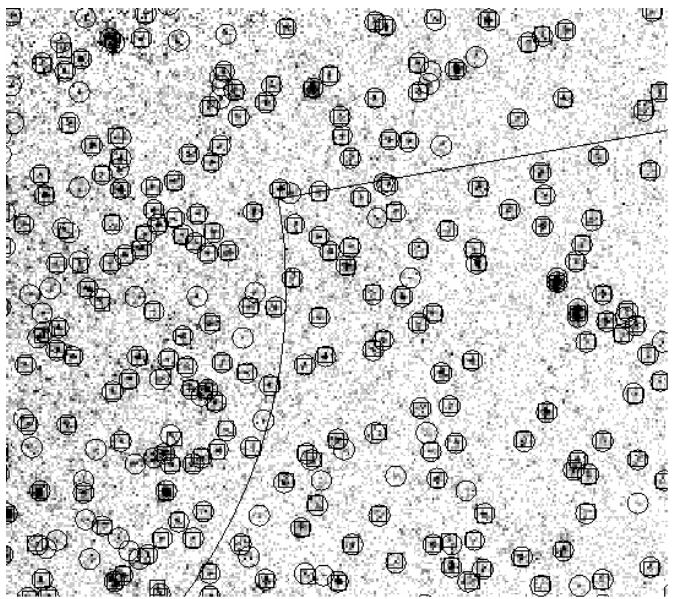}
\caption{Comparison of source detections by wavelet decomposition (circles)
  and the CIAO task \texttt{wavdetect} (boxes). Most of the sources are
  found by both detection algorithms.} 
\label{wvdecomp_and_wav}
\end{figure}

\begin{figure*}
\includegraphics[width=\textwidth]{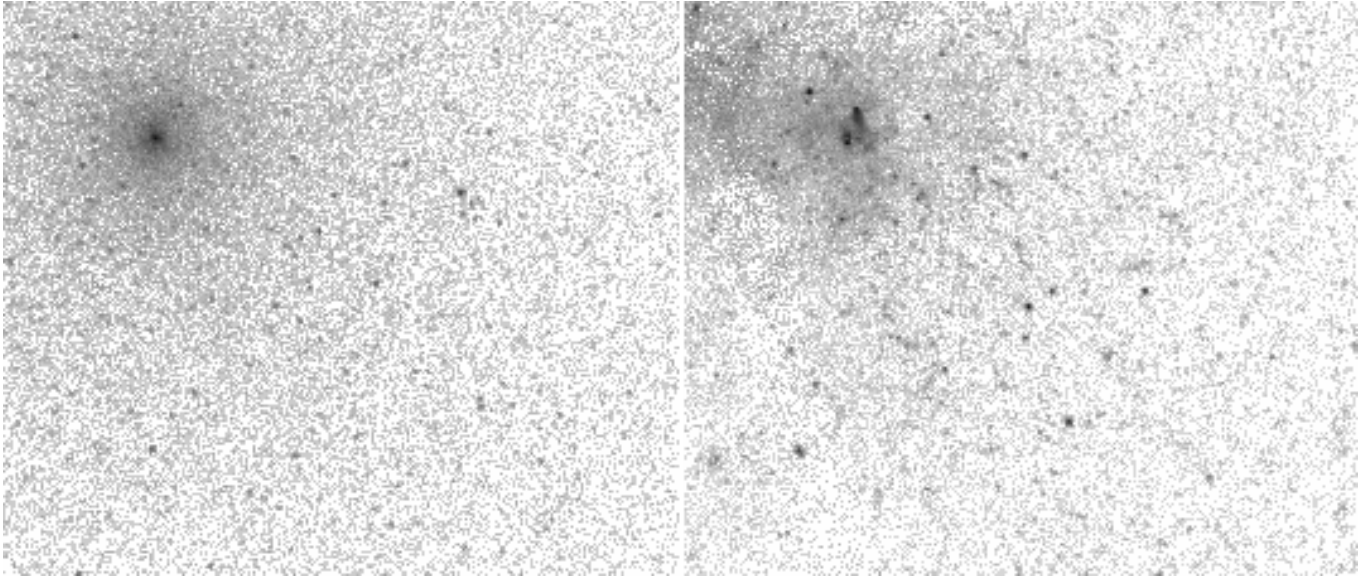}
\caption{Simulated image of the Galactic Center region (left) vs. the
real image taken by Chandra (right). One can see a close similarity
between them. The Galactic Center is in the upper left corner.} 
\label{twoimages}
\end{figure*}

To maximize the sensitivity to point sources, we used only the data
within $R<4\arcmin$ from the optical axis, where the \emph{Chandra}
angular resolution is within $1''$ (FWHM). Using a stacked image of
bright ($>100$~cnts) sources, we verified that the effective Point
Spread Function (PSF) in the combined image can be well modeled as a
Gaussian with $\sigma\sim0.51\arcsec$.

\section{Source counting strategy}

To determine the total point source flux we cannot simply coadd the
observed counts from detected sources, because of strong statistical
biases discussed below. A better approach is to integrate the
reconstructed $\log N - \log S$ function down to the sensitivity limit.

Because of its excellent angular resolution and low instrumental
background, \emph{Chandra} can detect point sources yielding just a few
total counts. Such low detection thresholds lead to strong biases in the
derived $\log N - \log S$ related to the Poisson counting statistics.
Generally, the raw number-flux relation near the detection threshold is
significanly below the true $\log N - \log S$ because of these effects.
This observational bias and methods for correcting it have been
extensively studied \citep[e.g.][ and references
therein]{hasinger93,vikhlinin95}, and in particular for the Chandra
image analysis in \cite{moretti02,kenter03,bauer04}. An additional bias
in our case arises from source confusion, non-neglibile in the GC region
even with the \emph{Chandra}'s angular resolution (Fig.\ref{image_sources}).

Our procedure for recovering the true $\log N - \log S$ function in the
GC region is based on Monte-Carlo simulations, which allows us to derive
accurate corrections for the statistical and source confusion biases. We
start with simulating the images containing realistic parent point
source populations. We assume that the shape of the $\log N - \log S$
relation is independent of distance from Sgr~A$^*$ (which corresponds to
distance-independent luminosity function in this region). The
normalization of the parent $\log N - \log S$ is assumed to vary as
$dN/d\Omega\propto R^{-1}$, consisted with the previous studies of both
the descrete source populations \cite[see e.g.][]{muno03,muno06} and the
total X-ray flux \cite[e.g.][]{neronov05} in the GC region. The shape of
the parent $\log N - \log S$ is assumed to be a power law, $dN/dS\propto
S^{-\alpha}$ with $\alpha=3.0$, at bright fluxes
($f>40$~cnt)\footnote{Assuming the power law spectrum with the photon
  index $\Gamma=2$ absorbed with $N_{\rm H}\sim 5\times
  10^{22}$~cm$^{-2}$ \citep{muno04}, 40 counts in the 4--8~keV band
  corresponds to an unabsorved flux of
  $7\times10^{-15}$~erg~s$^{-1}$~cm$^{-2}$ in the 2--10~keV band.}. At lower fluxes, we
allowed for a nearly arbitrary shape of the intrinsic $\log N - \log S$.
Namely, we assumed that $S^2dN/dS={\rm const}$ in the flux intervals
40--10, 10--3, 3--1, and 1--0.3 cnts but the normalization within each
interval is arbitrary. The normalizations of the parent $\log N - \log
S$ in each flux range were varied randomly in each realization. For each
simulated source, the number of detected photons was drawn from the
Poisson distribution. If the total number of simulated photons was below
the total observed intensity, the ``missing'' flux was added as a
diffuse component with the surface brightness $\propto R^{-1}$. An
example of the simulated image is shown in Fig.\ref{twoimages}.

Our source detection is based on the wavelet decomposition algorithm
described in \cite{vikhlinin98}. The detection threshold (specified as a
required statistical significance) was chosen so that less than 1 spurious
detection is allowed over the region of interest. Sources detected in the
\emph{Chandra} image of the GC region are shown by circles in
Fig.\ref{image_sources}. We also checked that wavelet decomposition provides
equal or better sensitivity to the point sources compared with the CIAO task
\texttt{wavdetect} (Fig.~\ref{wvdecomp_and_wav}).

The fluxes of detected sources were measured within the $2.5''$ aperture
(which should contain nearly 100\% of the total source flux for the observed
PSF width, see \S\,\ref{sec:data-reduction}). The local background, provided
by the largest scale of the wavelet decomposition, $\sim 16''$, was
subtracted from the source flux.

\begin{figure}
\includegraphics[height=\columnwidth,bb=60 150 503 720,clip,angle=-90]{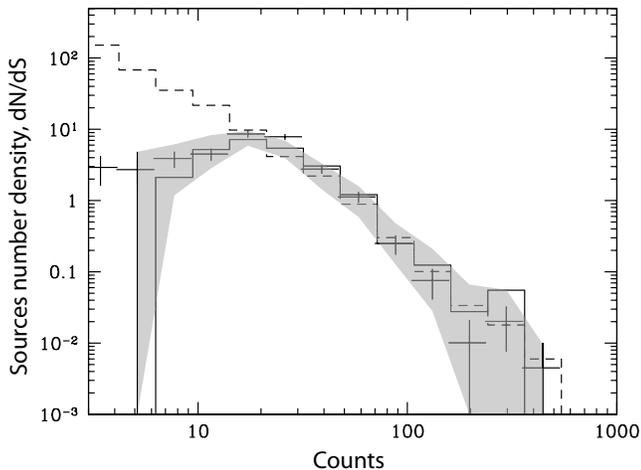}
\caption{Differential number-flux function of sources in the region of
our study (points with error bars) along with a well-matching
simulated number-flux distribution (solid line) and the corresponding 
parent (before the detection procedure) distribution of sources
(dashed line). Gray area denotes the manifold of trial shapes 
of differential luminosity functions of simulated sources 
which we considered as satisfactorily 
describing the observed $dN/dS$ function} 
\label{diff_counts}
\end{figure}

The raw differential number-flux relation derived from the real
observation is shown by crosses in Fig.\ref{diff_counts}. From our
simulations, we can determine the ``response'' to which shape and
normalization of the parent $\log N - \log S$ best describes the
observed flux distribution. These results are discussed below.

\section{Results}

The region outlined by the solid line in Fig.~\ref{image} contains in total
63.5~kcnts in the energy band 4--8 keV. The estimated particle background
image of the same region contains 10.6 kcnts. Thus the net
background-corrected number of counts is 52.9~kcnts. The total flux from all
the sources detected in the region is $\sim 14.3$ kcnts.

\subsection{Resolved fraction}

Figure~\ref{diff_counts} shows an example of a simulated number-flux
function that is consistent with the distribution measured by {\em Chandra}.
Also shown is the corresponding intrinsic $\log N - \log S$ relation. By
sampling over many such well-matching (the reduced $\chi^2$ difference from
the observed number-flux function is less than 1.5) trial distributions, we
determined the allowed range of the cumulative flux of ``parent'' sources in
the GC region (shaded region in Fig.~\ref{diff_counts}).

\begin{figure}[htb]
\includegraphics[width=\columnwidth,bb=28 150 570 720,clip]{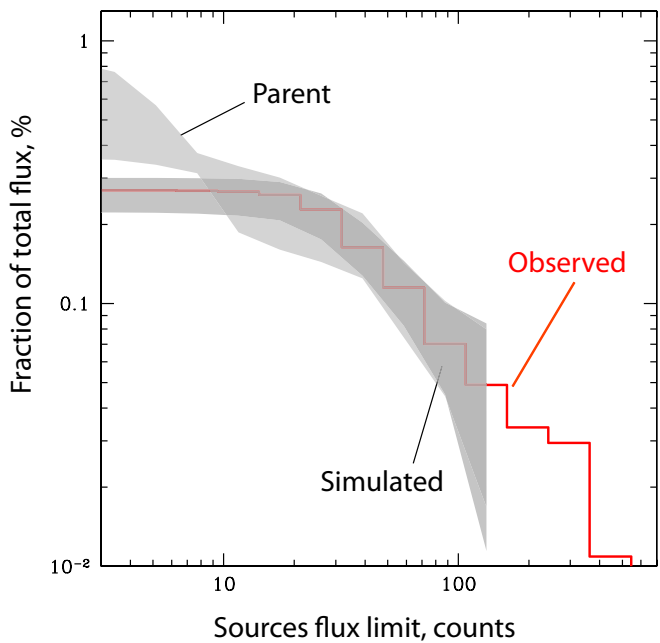} 
\caption{Cumulative flux (relative to the total flux) from
the detected sources in the real and simulated observations and the
allowed range for the parent source distribution. Note that one count
during the whole observation corresponds to an intrinsic (corrected
for the line-of-sight extinction of $N_{\rm H}=5\times10^{22}$ cm$^{-2}$)
source luminosity $\sim 1.3\times10^{30}$ \lum.} 
\label{cumflux}
\end{figure}

The outcome of this analysis is presented in Fig.~\ref{cumflux}. We
conclude that at least 40\% and possibly 100\% of the total 
X-ray flux from the GC region is produced by point sources. We note
that in obtaining this result we limited ourselves to sources with
average intensities of more than 3 integrated counts,
because the number densities of yet weaker sources are 
poorly constrained by the available data. Therefore 40\% is actually a
conservative lower limit on the contribution of point sources to the
total X-ray flux. Below we discuss the expected contribution of sources with
luminosities below the Chandra detection threshold. 

\subsection{Luminosity function of point X-ray sources}

We now address the luminosity function of faint X-ray sources
in the GC region. Since the volume density of stars within several tens
of parsecs of Sgr A$^*$ is several orders of magnitude higher than in
other places along the line of sight in that direction, we can safely
assume that the absolute majority of stars and X-ray sources in the
sky region of our study (located at a $\sim 10$~pc projected distance
from Sgr A$^*$, see Fig.~\ref{image}) physically belong 
to either the nuclear stellar cluster (NSC) or nuclear stellar disk (NSD)
components of the Galaxy \citep{genzel87,launhardt02}. This allows us
to readily estimate the stellar mass contained in the volume of 
the Galaxy covered by our observations and consequently the luminosity
function of X-ray sources in the GC region normalized by stellar mass. 

For the NSC we adopted the following density profile:
$$
\rho_{\rm NSC}={\rho_c\over{1+(r/r_c)^n}},
$$
where $\rho_c=3.3\times10^6 M_\odot$/pc$^3$ and $r_c=0.22$~pc. The
slope was assumed to be $n=2$ at $r<6$ pc and  $n=3$ at larger
distances. The total mass of the NSC within 200 pc of Sgr A$^*$ is
thus $6\times10^{7} M_\odot$. This value is actually uncertain by a
factor of $\sim 2$ \citep[see e.g.][]{lindqvist92,launhardt02}. 

The NSD was assumed to have the density distribution ($r$ and $z$ are
measured in parsecs)
$$
\rho_{\rm NSD}=\rho_d r^{-\alpha}e^{-|z|/z_d},
$$
where $\rho_d=300 M_\odot$/pc$^3$ and $z_d=45$ pc. At $r<120$ pc, the slope
$\alpha=0.1$, at $120$ pc $<r<220$ pc, $\alpha=3.5$, and at $r>220$ pc,
$\alpha=10$. The total mass of the NSD is thus $1.4\times10^9
M_\odot$. In reality this quantity is uncertain by some 50\%
\citep{launhardt02}. We took the distance to the Galactic Center to
be 7.6 kpc \citep{eisenhauer05}. 

The adopted mass model gives a total mass of stars enclosed in the
region of our study of $\sim 4.4 \times 10^{6} M_\odot$. This value is
uncertain by a factor of $\sim 2$ mainly due to the uncertainty in the NSC
mass. We note that a similar mass model was found to provide a
satisfactory description of the surface density distribution of
detected X-ray sources in the GC region \citep{muno06}.   

\begin{figure}[htb]
\includegraphics[width=\columnwidth,bb=10 140 558 730,clip,angle=-90]{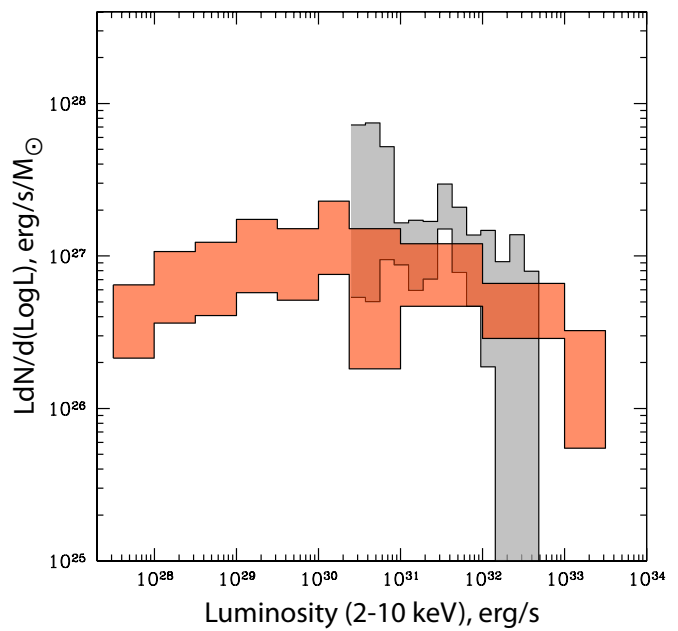} 
\caption{Luminosity function of weak X-ray sources in the Solar neighborhood 
(red area) in comparison with the allowed range for the luminosity
function in the GC region (grey area). The points with error bars show
a luminosity function estimated directly from the GC 
number-flux function, without correcting for the bias
associated with Poisson counting noise at low fluxes. The 
normalization of the GC luminosity function per unit stellar mass is
uncertain by a factor 
of $\sim 2$ due to the uncertainty in the mass of the nuclear stellar cluster.} 
\label{nufnu}
\end{figure}

Using this estimate of the enclosed stellar mass and the Chandra
source number-flux function, we obtained from simulations the allowed
range for the luminosity function of X-ray sources in the GC
region at luminosities $L_{\rm 2-10~keV}> 10^{30}$ \lum, which is shown in
Fig.~\ref{nufnu}. It is important to note that the overall
normalization of this luminosity function is uncertain by a factor of
$\sim 2$ due to the uncertainty in the stellar mass enclosed in the
studied region of the Galaxy. For comparison we show in
Fig.~\ref{nufnu} (points with error bars) the result of simply
dividing the differential $\log N$--$\log S$ function measured by
Chandra by the enclosed stellar mass. This naive determination is
affected by the bias discussed in the previous section and
thus underestimates the true luminosity function. On the other hand,
the allowed range for the luminosity function inferred from the
Chandra data for the dense GC region is remarkably compatible with the
luminosity function of faint X-ray sources in the Solar vicinity 
(\citealt{sazonov06}, Fig.~\ref{nufnu}), taking into account the
uncertainty in the NSC mass.  

The total absorption-corrected (assuming a line-of-sight absorption
of $N_{\rm H}=5\times 10^{22}$~cm$^{-2}$) X-ray flux in the
energy band 2--10~keV from the studied GC region is $F_{\rm
x}=(5.8\pm0.4)\times10^{-12}$ \flux. For the GC distance of 7.6 kpc
\citep{eisenhauer05}, this corresponds to a total X-ray luminosity
$L_{\rm 2-10~keV}=(4.0\pm0.2) \times 10^{34}$ \lum. Therefore the
total X-ray emissivity in the studied volume of the 
Galaxy is $L_{\rm x}/M=(9.1\pm 4.6)\times 10^{27}$ 
\lum $M_\odot^{-1}$, where we included the uncertainty in the NSC
mass. This derived value of the X-ray emissivity
per unit stellar mass agrees within the uncertainties with the
cumulative emissivity of faint X-ray sources (cataclysmic variables
and coronally active stars) near the Sun:
$(4.5\pm0.9)\times 10^{27}$ \lum $M_\odot^{-1}$
\citep{sazonov06}. 

We should note here that in the studied GC region there may be a non-negligible
contribution from warm diffuse plasma heated by supernova
remnants. According to \cite{muno04}, $\sim
15$\% of the total 2--10 keV flux is probably due to a $\sim
0.8$ keV plasma. This would mean that the X-ray emissivity of point
sources in the GC region is actually somewhat smaller, $L_{\rm
x}/M=(7.7\pm 3.9)\times 10^{27}$ \lum $M_\odot^{-1}$, i.e. even closer
to the value measured by \cite{sazonov06} in the Solar neighborhood. 

\section{Discussion}
We showed that at least 40\% of the total X-ray emission from the
Galactic Center region is produced by point sources with
luminosities $L_{\rm 2-10~keV}> 10^{31}$ \lum. The inferred
luminosity function of such sources \citep[see also][]{muno06} is
compatible with that measured in the Solar vicinity
\citep{sazonov06}. Moreover, the data are consistent with the
hypothesis that sources with luminosities below $\sim 10^{31}$
\lum, the effective Chandra detection limit, 
provide the rest of the total X-ray flux. Such sources are in fact
expected to be present in the required numbers in the GC region. 

According to studies in the Solar neighborhood \citep{sazonov06}, most
of the X-ray sources with luminosities $L_{\rm 2-10~keV}<10^{31}$
\lum\ are coronally active stars. Late-type stars with convective
envelopes can have coronae where the plasma can be heated up to X-ray
temperatures \cite[see][for a review]{gudel04}. Such stars are usually
fast rotators, as is required for high coronal activity, because they
are either members of binary systems or relatively young
\citep[see e.g.][]{walter81,queloz98}. The relative fraction of binary
stars with active coronae near the Sun is quite high: at least every
80'th star has an X-ray (2--10 keV) luminosity in the range $10^{27}$~\lum\  
$<L_{\rm 2-10~keV}<2\times 10^{30}$ \lum. The relative fraction of single
stars with active coronae is even larger. 

The relative contribution of coronally active stars to the total X-ray
emission of a given volume of the Galaxy is estimated to be 30--60\%
\citep{sazonov06}. For this fraction to be smaller, the fraction 
of stars with convective envelopes (usually late-type, low-mass stars)
should somehow be reduced, whereas the population of low-mass stars at
distances 4.5--8.7 pc from Sgr A$^*$ appears quite normal
\citep[e.g.][]{philipp99}.  

As mentioned above, the luminosity function of GC sources agrees
within the uncertainties with that measured in the Solar 
vicinity \citep{sazonov06}. We point out that because the vast 
majority of sources detected in the GC region are located at the same
and known distance from us, a luminosity function determined there
can be more accurate than a luminosity function determined
elsewhere.
 
\begin{acknowledgements}

This research made use of data obtained from the High Energy Astrophysics
Science Archive Research Center Online Service, provided by the
NASA/Goddard Space Flight Center. 
\end{acknowledgements}

\end{document}